\newcommand{\CLASSINPUTinnersidemargin}{18mm}
\newcommand{\CLASSINPUToutersidemargin}{12mm}
\newcommand{\CLASSINPUTtoptextmargin}{20mm}
\newcommand{\CLASSINPUTbottomtextmargin}{25mm}

\documentclass[10pt,conference,a4paper]{IEEEtran}
\usepackage[utf8]{inputenc}

\usepackage{times}
\usepackage{color}

\newcommand*\justify{%
  \hyphenchar\font=`\-% allowing hyphenation
}
\let\textttOld\texttt %To make a local copy of the previous texttt in order to use it in the renewcommand
\renewcommand{\texttt}[1]{\textttOld{\justify #1}}

\usepackage{graphicx}
\usepackage{subfigure}
\DeclareGraphicsExtensions{.png,.eps,.ps,.pdf}

\usepackage{url}

\begin{document}

%%% Título
\title{MSNM-Sensor: An Applied Network Monitoring Tool for Anomaly Detection in Complex Networks and Systems}

%%% Autores
\author{
	\IEEEauthorblockN{\footnotesize{Roberto Magán Carrión}}
	\IEEEauthorblockA{\footnotesize{Dpt. of Computer Engineering} \\
		\footnotesize{University of Cádiz}\\
		\footnotesize{Network Engineering}\\
		\footnotesize{\&}\\
		\footnotesize{Security Group}\\
		\footnotesize{University of Granada}\\
		\footnotesize{\{roberto.magan@uca.es, rmagan@ugr.es\}}}
	\and
	\IEEEauthorblockN{\footnotesize{José Camacho}}
	\IEEEauthorblockA{\footnotesize{Dpt. of Signal Theory,} \\ \footnotesize{Telematics \& Communications,} \\
		\footnotesize{Network Engineering}\\
		\footnotesize{\&}\\
		\footnotesize{Security Group}\\
		\footnotesize{University of Granada}\\
		\footnotesize{\{josecamacho@ugr.es\}}}
	\and
	\IEEEauthorblockN{\footnotesize{Gabril Maciá-Fernández}}
	\IEEEauthorblockA{\footnotesize{Dpt. of Signal Theory,} \\ \footnotesize{Telematics \& Communications,} \\
		\footnotesize{Network Engineering}\\
		\footnotesize{\&}\\
		\footnotesize{Security Group}\\
		\footnotesize{University of Granada}\\
		\footnotesize{\{josecamacho@ugr.es\}}}
	\and
	\IEEEauthorblockN{\footnotesize{Angel Ruíz-Zafra}}
	\IEEEauthorblockA{\footnotesize{Dpt. of Computer Engineering} \\
		\footnotesize{University of Cádiz}\\
		\footnotesize{\{angel.ruiz@uca.es\}}}
}

\maketitle

%%% Abstract
\begin{abstract}
Technology evolves quickly. Low-cost and ready-to-connect devices are designed to provide new services and applications. Smart grids or smart healthcare systems are some examples of these applications, all of which are in the context of smart cities. In this total-connectivity scenario, some security issues arise since the larger the number of connected devices is, the greater the surface attack dimension. In this way, new solutions for monitoring and detecting security events are needed to address new challenges brought about by this scenario, among others, the large number of devices to monitor, the large amount of data to manage and the real-time requirement to provide quick security event detection and, consequently, quick response to attacks. In this work, a practical and ready-to-use tool for monitoring and detecting security events in these environments is developed and introduced. The tool is based on the Multivariate Statistical Network Monitoring (MSNM) methodology for monitoring and anomaly detection and we call it MSNM-Sensor. Although it is in its early development stages, experimental results based on the detection of well-known attacks in hierarchical network systems prove the suitability of this tool for more complex scenarios, such as those found in smart cities or IoT ecosystems.
\end{abstract}

%%% Palabras clave

\section{Introduction}
\label{sec:introduction}
Several technical reports forecast 30 billion IoT (Internet of Things) devices around the world by 2021 \cite{iot_devices_outdated} and  more than 3 billion M2M (Machine to Machine) connections by 2022 \cite{m2m_connections}. This scenario enables new services and applications for improving people's daily life as well as business opportunities. However, many challenges arise, with security being one of the most important. 

Underlying systems and communications networks are continually being threatened by attackers, even more in these hiper-connected environments. For instance, it is worth mentioning two recent attacks, Dyn (2016)~\cite{enisa_etl_2017,chaabouni_network_2019} and VPNFilter (2018)~\cite{enisa_etl_2018}, where thousands of Internet of Things (IoT) devices were compromised causing, on the one hand, a high economic impact and, on the other hand, and even worse, personal costs.

The way to monitor and control what is happening in these kinds of scenarios is a great challenge since the attack exposure surface grows almost exponentially with the number of devices interconnected. A more challenging issue is to manage the generated data gathered from different information sources such as applications, networking devices and communications. In this way, key aspects such as managing the volume, veracity or velocity of the data are crucial for achieving quick and efficient detection and reaction against security attacks \cite{camacho_tackling_2014}. Furthermore, these aspects may limit the practical application of the solutions, especially in the described scenario.

To address the previous issues, IDS (Intrusion Detection System) and SIEM (System Information and Event Management) tools are widely used by the specialized community on ICT (Information and Communication Technologies) security. IDS systems are commonly categorized as one of two types: 1) Network-based IDSs (NIDSs) and 2) Host-based IDSs (HIDSs). NIDSs monitor network events such as traffic flows or firewall logs, while HIDSs behave similarly but consider host (endpoint)-related events, \textit{e.g.}, syslog, CPU load, etc. Along with this categorization, we can also discern between Anomaly-based IDSs (AIDSs) and Signature-based IDSs (SIDSs), depending on the detection technique applied. The first one (AIDS) tries to find anomalous patterns in the monitored information, while the second one (SIDS) intends to detect the presence of an attack by comparing the gathered information with previously defined attack signatures. Snort \cite{snort_ids} and Zeek (Bro) \cite{bro_ids} are some examples of practical and widely used IDS tools.

SIEM tools also include monitoring and detection functionalities but also provide services for event correlation and reporting capabilities, among others. Moreover, SIEM tools can also handle information gathered from IDS systems as a suitable data source for alerting security events. Splunk \cite{splunk_siem} and AlientVault \cite{alient_siem} are some examples of practical and widely used SIEM tools.

However, neither IDSs nor SIEM tools can efficiently manage new monitoring and security event detection needs arising from new communication and computation paradigms. In this work the MSNM-Sensor (Multivariate Statistical Network Monitoring Sensor) tool is introduced to address the previous issues. It is based on the MSNM (Multivariate Statistical Network Monitoring) approach coined by Camacho \textit{et al.} \cite{camacho_pca-based_2016} and successfully tested afterwards in works \cite{camacho_traffic_2017, camacho_semi-supervised_2019,camacho_multivariate_2019}. Briefly, MSNM methodology comprises four main steps which are \textit{parsing}, \textit{fusion}, \textit{detection} and \textit{diagnosis}. All the previous stages has successfully been integrated into the MSNM-Sensor having a useful tool for monitoring and anomaly detection in real-time. Among others characteristics, it is able to manage heterogeneous data sources in real time; reduces the monitoring network traffic without significant impact in the detection performance; is lightweight, scalable, versatile and dynamically adaptable to changes in the network environments; keeps privacy on communications; provides a friendly interactive dashboard and it is an open source project.
	
In order to demonstrate its suitability to be used in complex network environments, MSNM-Sensor has been tested to monitor hierarchical networks and systems for detecting well-known attacks like \textit{DoS}, \textit{port scanning} and \textit{data exfiltration} as can be seen in \ref{sec:experimentation} section.

%\begin{enumerate}
%    \item Management of different and heterogeneous data sources in real time.
%	\item Greatly reduce the network monitoring traffic, but maintaining the detection performance. Against current IDS- and SIEM-based solutions fed with raw and heavy data, MSNM-S just only considers two statistics as monitoring information to monitor the whole system. It is a key point in complex network architectures and systems such as IoT ecosystems with a large number of devices to monitor. 
%	\item Lightweight, scalable, versatile, distributed and dynamically adaptable to changes in the system under monitoring, which is a relevant issue when addressing network communications.
%	\item Privacy in monitoring communications, because no raw data is sent to the central control.
%	\item An easy-to-use front end with an interactive dashboard to control the whole system.
%	\item An open-source project.
%\end{enumerate} 

The paper has been organized as follows. Section \ref{sec:msnm_fundamentals} describes the fundamentals of MSNM methodology, which supports the core functionality of the sensor. In Section \ref{sec:components_operations} the components and operation modes of MSNM-Sensor are described. The detection capabilities of the sensor are valida\-ted in realistic network architectures through  Section \ref{sec:experimentation}. Moreover, attacks such as DoS, data exfiltration or port scanning are successfully detected and located in the proposed network architecture. Finally, conclusions and further work are described in Section \ref{sec:conclusions}.

\section{Fundamentals of Multivariate Statistical Network Monitoring}
\label{sec:msnm_fundamentals}
In this section we briefly introduce MSNM, which is the basis of the MSNM-Sensor tool. MSNM relies on PCA (Principal Component Analysis), a main tool for multivariate analysis. PCA has been established as a promising technology to perform network anomaly detection \cite{kanaoka_multivariate_2003,lakhina_diagnosing_2004,callegari_improving_2014,delimargas_evaluating_2014}. Besides, the unsupervised nature of PCA allows to unveil anomalous behavior from unknown attacks which is a desired characteristic in those solutions working in real environments. Apart from its unsupervised nature, PCA is a white-box model. In comparison to the bulk of machine learning solutions, PCA models are explainable since trends or relationships in the features and observations of the data, can be easily connected. 

MSNM has been demonstrated to be a promising methodology in network anomaly detection through several works  \cite{camacho_pca-based_2016,camacho_traffic_2017,camacho_semi-supervised_2019,camacho_multivariate_2019,magan-carrion_multivariate_2015} offering a high detection performance in this context. Four stages comprise the MSNM methodology. They are: a) \textit{parsing}, b) \textit{fusion}, c) \textit{detection} and d) \textit{diagnosis}. All of them are introduced in the following and integrated in the MSNM-Sensor tool as can be seen in next section.

\subsection{Parsing}

Information from communication networks usually comes in the shape of huge logs and traffic based files containing heterogeneous information. This makes it impossible to directly use this information as input of detection systems and learning models to identify different kinds of attacks. However, to overcome this issue, data sources are processed in order to build a more suitable input for automatic detectors or classifiers.

In this sense, the application of some feature engineering technique is proposed to build a well-structured input, suitable for monitoring and detection systems in general. Thus, the Feature as a Counter (FaaC)~\cite{camacho_tackling_2014} technique is used as a functional solution to the problem of learning from large heterogeneous datasets. It consists in transforming different data sources (structured or not) of information with features into new variables. The new ones are just counters of the original ones computed in a given interval of time. For instance, it could be interesting to count the number of accesses to port 22 in a given time window interval, because a high number might mean a brute force SSH attack. The practical implementation of the FaaC approach, named FCParser, is available online for downloading at \cite{faac_github}.

\subsection{Fusion}
In communication networks and systems it is expected to find more than one information data source to be monitored. Apart from the \textit{parsing} functionality of FaaC, the FCParser tool is also able to \textit{fuse} different data sources in a single set of features. For each different source of data, a set of features (counters) is defined. All the sources under monitoring are appended into a simple data stream to build the calibration matrix for the PCA model. For that, each source is sampled with equal sampling rate, then parsed and finally fused into an unique data stream. This procedure is periodically repeated at each sample time.

The combination of the \textit{parsing} and the \textit{fusion} procedures is specially suited for the subsequent multivariate analysis, resulting in high dimensional feature vectors that need to be analyzed with dimension reduction techniques, like PCA. Moreover, the diagnosis procedure benefits from the definition of a large number of features for a better description of the anomaly taking place. Finally, counters and their correlation are easy to interpret.

\subsection{Detection}

PCA is the core of MSNM. PCA identifies a number of linear combinations of the original variables in a data set $\mathbf{X}$\footnote{$\mathbf{X}$ contains the data matrix obtained in the previous two stages.}, the so-called PCs (Principal Components), containing most of its relevant information (variability). This process involves a change in variables from the original ones in the $\mathbf{X}$ space to those in the PC subspace. If $\mathbf{X}$ contains $M$ variables and $N$ observations of each variable, PCA reduces its dimensions from $M$ variables to $A$ PCs by finding the $A$-dimensional latent subspace of the most variability captured.

PCA is described through the following equation:

\begin{equation}
\mathbf{X}=\mathbf{T}_{A}\cdot\mathbf{P}_{A}^{t}+\mathbf{E}_{A}
\label{eq:pca_model}
\end{equation}

\noindent where $\mathbf{P}_{A}$ is the $M\times$$A$ loading matrix, $\mathbf{T}_{A}$ is the $N\times$$A$ score matrix and $\mathbf{E}_{A}$ is the $N\times$$M$ residual matrix. The maximum variance directions are obtained from the eigenvectors of $\mathbf{X}^t\cdot\mathbf{X}$, and they are ordered as the columns of $\mathbf{P}_{A}$ by the explained variance. The rows of $\mathbf{T}_{A}$ are the projections of the original $N$ observations in the new latent subspace. $\mathbf{E}_{A}$ is the matrix that contains the residual error, and it plays a key role in the anomaly detection, as shown later.
The projection (score) onto the PCA subspace of a new observation is obtained as follows:

\begin{equation}
\mathbf{t}_{new}=\mathbf{x}_{new}\cdot\mathbf{P}_{A}
\label{eq:pca_obtencion_t}
\end{equation}

\noindent where $\mathbf{x}_{new}$ is a $1\times$$M$ vector that represents a new object and $\mathbf{t}_{new}$ is a $1\times$$A$ vector that represents the projection of the latent subspace, while

\begin{equation}
\mathbf{e}_{new}=\mathbf{x}_{new} - \mathbf{t}_{new}\cdot\mathbf{P}_{A}^{t}
\label{eq:pca_e}
\end{equation}

\noindent corresponds to the residuals.

In order to detect anomalies in the monitored system $Q$-st \cite{jackson_control_1979} and $D$-st (also known as $T^2$) \cite{hotelling1947} statistics are commonly used. $Q$-st compresses the residuals in each observation, and $D$-st is computed from the scores. Their values for a specific observation can be computed through the following equations:

\begin{equation}
D_n=\sum_{a=1}^{A}\left({\displaystyle\frac{\tau_{an} - \mu_a}{\sigma_a}}\right)^2
\label{eq:t_cuadrado}
\end{equation}
\begin{equation}
Q_n=\sum_{m=1}^{M}\left(e_{nm}\right)^2
\label{eq:q_contribucion}
\end{equation}

\noindent where $\tau_{an}$ represents the score of the $n$-th observation of the $a$-th PC, $\mu_a$ and $\sigma_a$ stand for the mean and standard deviation for the scores of that PC in the calibration data, respectively, and $e_{nm}$ represents the residual value corresponding to the $n$-th observation of the $m$-th variable.

With both statistics (\ref{eq:t_cuadrado}) and (\ref{eq:q_contribucion}) computed from the calibration data $\mathbf{X}$, Upper Control Limits (UCLs), \textit{i.e.} detection thresholds, can be established with a certain confidence level \cite{camacho_pca-based_2016, nomikos_multivariate_1995}. Subsequently, new data are monitored using these limits, and an anomaly is identified when UCL limits are exceeded by new incoming observations. 
	
%Furthermore, the contribution of the variables to a detected anomaly can be investigated with the contribution plots \cite{kourti1996}, as discussed in Section \textit{\nameref{sec:experimentation}}.}

\subsection{Diagnosis}

After detecting something anomalous in the system, it is necessary to find out where the event came from and the reason as well. Most often, the diagnosis procedure is manually carried out by a security analyst alerted by the system.

The diagnosis procedure could be a tricky and tedious task, due to the large amount of information to analyze. Feature contributions to a given anomaly can be a useful tool to identify where the anomalous behavior comes from. Contribution plots or other diagnosis methods like oMEDA (observation-based Missing-data method for Exploratory Data Analysis) \cite{camacho_observation-based_2011} or Univariate Squared (US)~ \cite{nmfuentes_2018} can be used to identify the feature contributions. Thus, anomalies are detected in the $D$-st and/or $Q$-st statistics, and then the diagnosis is performed with \textit{e.g.}, oMEDA. For instance, The output of oMEDA is a $1 \times M$ vector where each element contains the contribution of the corresponding feature to the anomaly under study. Those contributions with large magnitude, either positive or negative, are considered to be relevant.

%Note that the optimum number of latent variables in the prediction sense does not necessarily match the optimum number for process monitoring \cite{camacho2012}.

\section{Multivariate Statistical Network Monitoring Sensor: MSNM-Sensor}
\label{sec:components_operations}
In the following section, MSNM-Sensor modules are introduced and thoroughly explained. In addition to others, like those in charge of inter-sensors communications or information sources management, the most important ones corresponds to the four steps found in MSNM methodology. Apart from that, the principal sensor operations are also described through illustrative examples.

\subsection{Modules}
The involved MSNM-Sensor functional modules, depicted in Figure \ref{fig:modules}, are described as follows:

\subsubsection{IS (Information Source)}
The IS module handles the data coming from the information sources. Two types of data sources, according to their location, are considered: 

\begin{itemize}
	\item Local (LIS). The information gathered from these data sources is generated by the device where the MSNM-Sensor is deployed. For instance, local information sources can be obtained from firewall log files, NetFlow traffic flows and host-based information (\textit{e.g.}, syslog in Linux-based systems), among others. LIS data is processed by the PARSING \& FUSION modules.
	\item Remote (RIS). The incoming data from other MSNM-Sensors are handled as a remote information source. Most of the data from other MSNM-Sensors will be the computed values of the monitoring statistics $Q_{n,m}$-st and $D_{n,m}$-st, allowing anomaly detection in complex and hierarchical systems. However, any other information could be obtained from a remote sensor, \textit{e.g.}, those requested by a security analyst for an anomaly diagnosis.
\end{itemize}

\subsubsection{PARSING \& FUSION }\label{subsec:parser_module}
As mentioned in Section \ref{sec:msnm_fundamentals} \textit{parsing} and \textit{fusion} modules do, on one hand, a transformation of LIS data sources (structures or not) into a new structured form where new features are counters of the original. On the other hand, in case of considering several LIS data sources, all of them are fused by appending one after another. As a result, we have a homogeneous data stream of quantitative values to be afterwards used in the DETECTION module for anomaly detection. Both processes are carried out by the FCParser \cite{faac_github} which implements the FaaC approach.

\subsubsection{DETECTION}\label{subsec:detection_module}

This module represents the sensor functional core for anomaly detection. It provides multivariate statistical-based methods and algorithms to compute the monitoring statistics $Q_{n,m}$-st and $D_{n,m}$-st in real time. This module enables the detection of anomalies in the monitored system when statistics computed from a new observation exceed certain control limits. Two control limits are calculated: UCLq and UCLd for $Q_{n,m}$-st and $D_{n,m}$-st respectively. Detailed information on monitoring statistics and UCL limits construction can be found in \cite{camacho_pca-based_2016}.

The DETECTION module supports three different functions:

\begin{itemize}
	\item \textbf{Preprocessing}. The DETECTION module performs data preprocessing for new observations and learning models. The standard normalization is used by default, but additional methods are also available and ready to use.
	\item \textbf{Modeling}. This operation is in charge of the sensor calibration from a set of observations that are, ideally, under NOCs (Normal Operation Conditions), \textit{i.e.}, without known anomalies. Currently, the PCA model is used, but other machine-learning-based techniques can be easily integrated \cite{bhuyan_network_2014}\cite{li_machine_2019}. 
	\item \textbf{Monitoring}. This operation computes the above mentioned statistics for each incoming observation. In addition to the precomputed control limits, the monitoring operation is able to detect anomalous behavior when these limits are exceeded. 
\end{itemize}

Because of the dynamic nature of information coming from networks and systems, learning models should periodically be re-calibrated or re-trained in order to adapt it to normal changes in the environment that avoids high false positive rates. Usually, such changes are due to the own cyclo-stationary nature of the information from network communications and systems \cite{macia-fernandez_ugr16_2018} which are different in volume and variety depending on the hour, day, week, etc. These behavioral patterns are periodically repeated. The EWMA (Exponentially Weighted Moving Average) approach is used to dynamically calibrate the sensors every 60 minutes \cite{camacho_visualizing_2014}. Each sample rate, new UCL limits are also computed.

\begin{figure}[t!]
	\begin{centering}
		\includegraphics[width=0.5\textwidth]{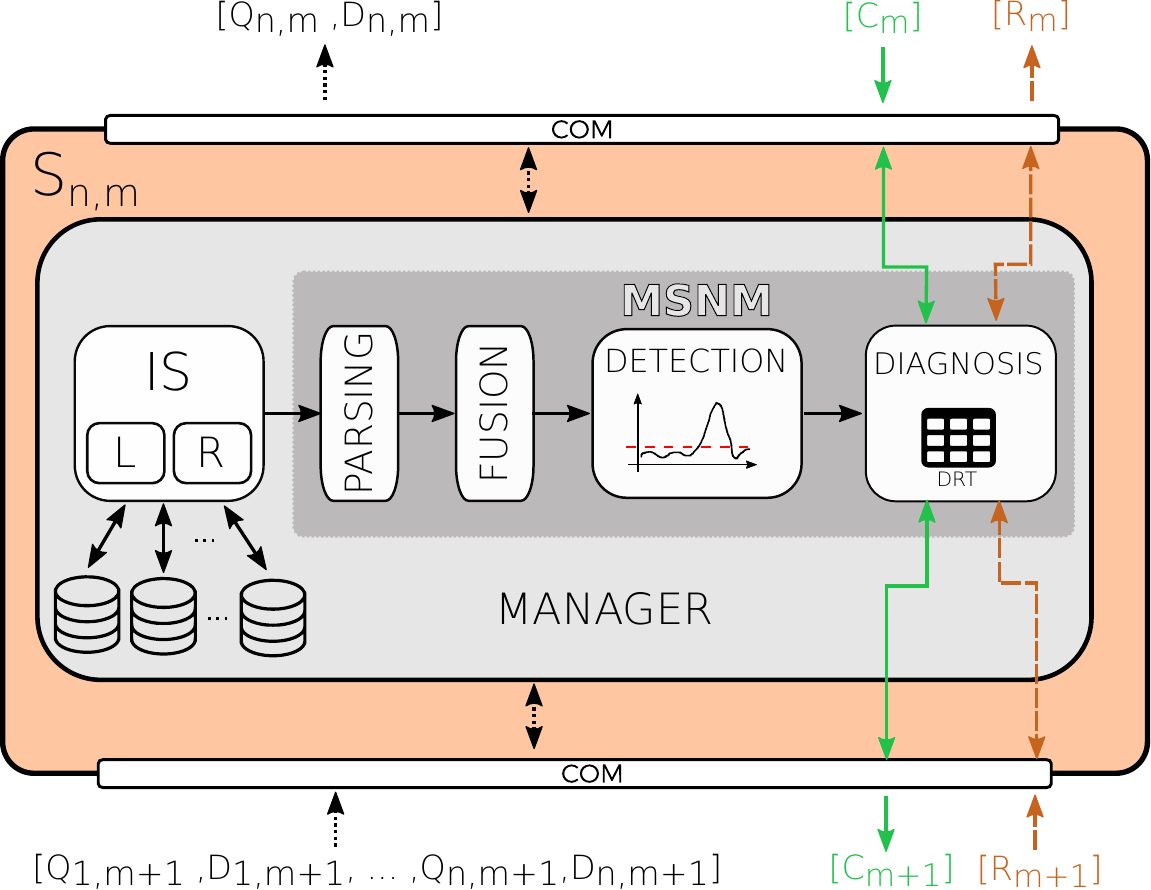}
		\par
	\end{centering}
	\caption{Functional modules of a generic sensor $S_{n,m}$. $n$ corresponds to the sensor ID, and $m$ is the hierarchical level ID where the sensor $n$ is deployed.}
	\label{fig:modules}
	
\end{figure}

\begin{figure*}[t!]
	\begin{centering}
		\includegraphics[width=1\textwidth]{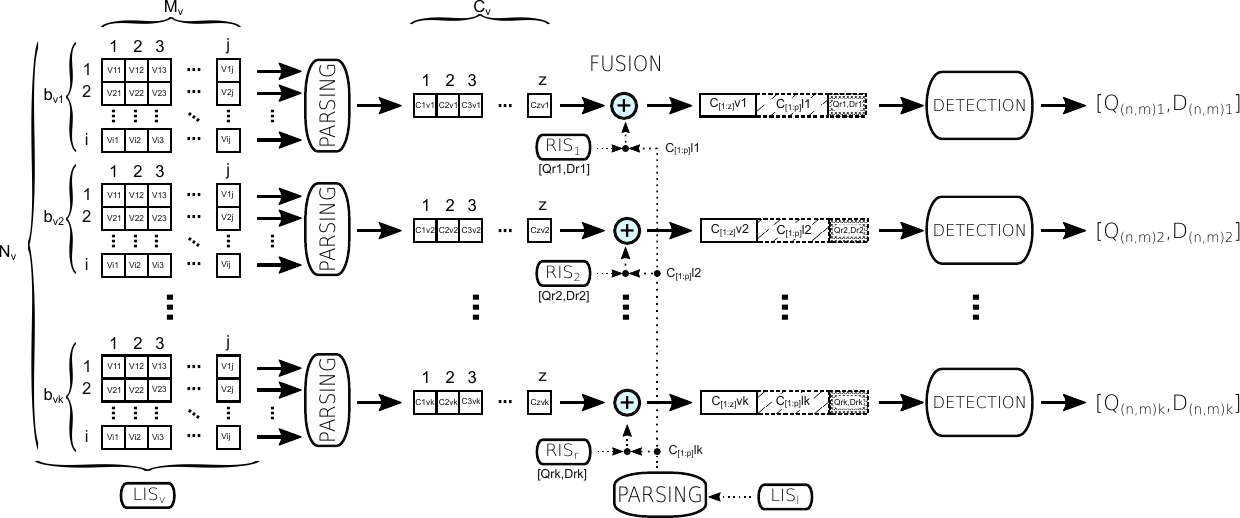}
		\par
	\end{centering}
	\caption{Involved modules and the information exchanged among them for an $S_{n,m}$ monitoring flow. The information comes from different data sources, namely, two local ($LIS_v$ and $LIS_l$) and one remote ($RIS_r$), to be divided into $k$ batches ($b_{v1},b_{v2},\dots,b_{vk}$). LIS sources (raw) are parsed for each batch. The new variables, counters of the original ones, are fused together with the remote ones. All variables together conform the observation to be monitored. After that, the $Q_{n,m}$ and $D_{n,m}$ statistics are computed by the MSNM module.}
	\label{fig:monitoring_operation_flow}
	
\end{figure*}

\subsubsection{DIAGNOSIS}

The diagnosis procedure takes place when an anomaly is detected on the monitored system. It is necessary to find out where the event came from and the reason as well, to afterwards deploy the adequate response measures to counteract against the attack. How to manage this problem is the duty of the DIAGNOSIS module.

The DIAGNOSIS module relies on the use of statistical multivariate techniques to determine the source of the anomaly. Currently, the oMEDA (observation-based Missing-data method for Exploratory Data Analysis) \cite{camacho_observation-based_2011} method is implemented, but other methods can be included, for instance, the contribution plots of the diagnosis method proposed in \cite{nmfuentes_2018}.

Although the DIAGNOSIS module solves the anomaly diagnostic by itself, it is done locally, \textit{i.e.}, in the device where the sensor was deployed. In addition, defining the device and data source(s) involved in an anomalous behavior in the whole system under monitoring is not a trivial task. For this reason, we create the DRT (Diagnosis Routing Table), which acts similarly to the well-known routing IP tables but adds together the information of local and remote sources. The diagnosis flow and routing procedure are explained in detail in Sections \ref{sec:sensor_operations} and \ref{sec:working_together}.

\subsubsection{COM (COMmunications)}

The COM module allows each MSNM-Sensor to exchange information. In this way, the COM module handles the exchange of specific messages. The system supports (but is not limited to) two types of messages: data and command. The messages mainly differ in the payload and type. For instance, a data message can include any information required in sensor operations, \textit{e.g.}, the monitoring statistics. On the other hand, command messages are devised to control these processes.

Depicted in Figure \ref{fig:modules}, two information flows are clearly differentiated: monitoring and diagnosis. It is worth mentioning that only the first one (monitoring) is currently implemented, while the second one is an ongoing work (see the project in \cite{msnm-sensor_github}). However, we decided to mention and describe both of them because they are complementary. In this way, monitoring information flow exchanges data messages containing the computed statistics $Q_{n,m}$ and $D_{n,m}$, while the diagnosis flow controls the entire diagnostic procedure.

In this early stage, there is no specific routing algorithm between sensors. Instead, each sensor must be manually configured to send and receive data to and from others. A self-deployed sensor process will be added in future releases.

Both flows and the exchanged messages are described in Sections \ref{sec:sensor_operations} and \ref{sec:working_together}.

\subsubsection{MANAGER}

As depicted in Figure \ref{fig:modules}, all modules work together following an execution pipeline and sharing the necessary information. The module in charge of orchestrating and managing the others is the MANAGER.

As mentioned above, there are two different information flows: monitoring and diagnosis. The first one (monitoring) involves four main modules (IS, PARSING, FUSION and DETECTION) that should be invoked sequentially, because the output of each module is the input of the next one. However, the second flow (diagnosis) involves the DIAGNOSIS module, which is invoked if a specific message is received. Finally, MANAGER handles the orchestration of which MSNM-Sensor module should run and when. 

A highly detailed description of the module interactions in each sensor is given in Section \ref{sec:sensor_operations}.

\subsection{Operations}
\label{sec:sensor_operations}

Thus far, the main functional MSNM-Sensor modules have been described. However, high-level operations, including several modules, are devised in accordance with the principal MSNM-Sensor functionalities: monitoring and diagnosis. The diagnosis process is still an ongoing work; however, we decide to briefly describe it for completeness. 

\subsubsection{Monitoring Operation}
To be aware of what is happening in systems or networks, it is important to detect anomalous behaviors (deliberate or not). However, this is not a trivial task, since a previous work must be done to select which element and information should be supervised. In this manner, the monitoring flow and the involved modules offer a versatile and scalable tool allowing the user to freely select data sources and variables to be monitored.

Figure \ref{fig:monitoring_operation_flow} shows a detailed view of module interactions and the exchanged information. In the figure, one can see a hypothetical local data source $LIS_v$ with $M_v$ variables to monitor a total of $N_v$ gathered observations, the latter being split into $k$ batches ($b_{v1},b_{v2},\dots,b_{vk}$). Each batch has $j=M_v$ variables and $i$ different numbers of observations each. As a result, we are able to 1) process less information, reducing the computation time, which is a key point for real-time applications, and 2)  adapt the monitoring time step granularity, sometimes hardly limited by the monitored data source or the anomaly to be detected. As explained in Section \ref{sec:experimentation}, 60 seconds is enough to monitor \textit{NetFlow}-based data sources in the detection of DoS attacks, for example.

For each $b_{vk}$ batch of observations, a new one is generated by parsing and fusing the original information (raw). This task is the duty of the PARSING \& FUSION modules (see Section \ref{subsec:parser_module}). At the time of writing this article, the implemented module was the FCParser \cite{faac_github} which implements the FaaC approach, a new feature engineering methodology that transforms the original information variable space into a new one where the new variables ($z$) are counters ($C_v$) of the original ones. This smart transformation makes the system highly versatile and scalable, allowing the user to define a large number of different new variables from a limited original set. For instance, counting the number of different destination ports in a certain period of time could be relevant for port scanning or port knocking attacks.

Although just one data source has been considered so far, additional local (LIS) or remote (RIS) data sources can also be included if needed. Figure \ref{fig:monitoring_operation_flow} shows the procedure to add new data sources. In this case (but not limited to it), there are three data sources involved: two local ($LIS_v$ and $LIS_l$) and one remote ($RIS_r$). At each monitoring step, a new fused observation is created. In this way, the extended space will have $e$ variables, with $e=z+p+2$, where $z$ is the number of variables (counters) from a batch $k$ of source $LIS_v$; $p$ is the number of variables (counters) from the same batch of source $LIS_l$; and $RIS_r$ has two variables as the number of statistics generated by the remote sensor. This observation is the input of the DETECTION module which is in charge of computing the monitoring statistics ($Q_{(n,m)k}, D_{(n,m)k}$). At this point, the system can detect anomalous behaviors when the control limits are exceeded. In addition, if this sensor is not the root in the hierarchy, the generated statistics will also be sent to the corresponding remote sensor for hierarchical monitoring and anomaly detection.

\begin{figure}[t!]
	\begin{centering}
		\includegraphics[width=0.5\textwidth]{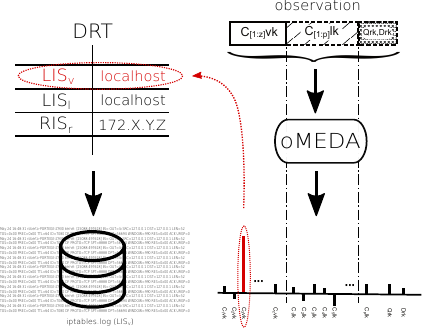}
		\par
	\end{centering}
	\caption{Diagnosis operation steps to determine the origin of an anomaly in the system. In this case, the anomaly comes from a variable ($C_3vk$) with a high contribution in the observation under inspection. This variable corresponds with a local data source $LIS_v$, which is a firewall log.}
	\label{fig:diagnosis_operation_flow}
	
\end{figure}

\subsubsection{Diagnosis Operation}
As aforementioned, after detecting an anomaly in the system, the diagnosis procedure should be launched to determine its origin. Figure \ref{fig:diagnosis_operation_flow} depicts the operation steps launched to determine which IS is involved in the anomaly. For instance, we suppose that a security event arose from the $LIS_v$ local data source that in turn represents an \textit{iptables} firewall.

Once the corresponding observation to be inspected has been retrieved from the sensor, the oMEDA algorithm is launched. This algorithm outputs the contribution of each variable, where those variables with high values (either positive or negative) must be inspected in detail as they are relevant. In this case, the variable with a high positive value is $C_3vk$ (the one in red in Figure \ref{fig:diagnosis_operation_flow}), which belongs to the $LIS_v$ data source according to the DRT. The DRT structure and role in this operation are described in the next section.

Together, the timestamp and the observation under investigation allow the sensor to extract the corresponding piece of raw information to be afterwards analyzed by the security analyst. 

\subsection{Working All Together: A Hierarchical Approach}
\label{sec:working_together}
As aforementioned, a single MSNM-Sensor instance is able to monitor and detect anomalous behaviors from a wide range of heterogeneous local and remote data sources. However, the really novel idea behind the use of $Q_{(n,m)}$ and $D_{(n,m)}$ statistics is their capability of maintaining the monitoring and anomaly detection performance when they are included in the hierarchy of complex network environments. This useful characteristic has been demonstrated in work \cite{macia-fernandez_hierarchical_2016}.

\begin{figure}[t!]
	\begin{centering}
		\includegraphics[width=0.5\textwidth]{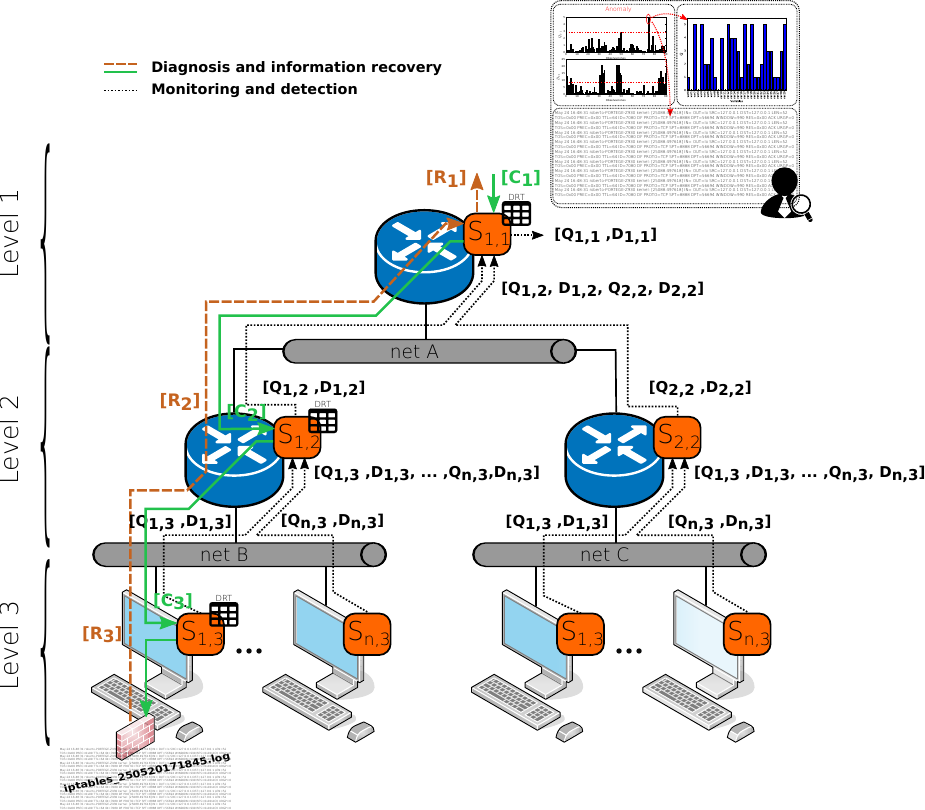}
		\par
	\end{centering}
	\caption{A hypothetical deployment of several MSNM-Sensors (orange boxes) throughout an interconnected system for hierarchical monitoring, anomaly detection (dashed black lines) and diagnosis (solid green and dashed brown lines).}
	\label{fig:working_together}
	
\end{figure}

Although the tool will be tested in Section \ref{sec:experimentation}, in a realistic network scenario, an example of several MSNM-Sensors cooperating within a hypothetical and common network deployment is described as follows for clarification. Figure \ref{fig:working_together} shows a simple network scenario where several MSNM-Sensors are deployed at hosts and network devices. We can discern in the figure two involved information flows: the monitoring and diagnosis flows. The former (black dashed lines) transports pairs of monitoring statistics ($[Q_{(n,m)}, D_{(n,m)}]$) coming from lower to higher levels in the hierarchy. In this synthetic example, sensors $S_{1,3},\cdots, S_{n,3}$ are deployed at hosts in the deepest architecture level, sending the generated statistics $[Q_{(1,3)}, D_{(1,3)}],\cdots,[Q_{(n,3)}, D_{(n,3)}]$ to the next closest sensor in the hierarchy. Indeed, they act as remote sources of $S_{1,2}$. Now, this sensor aggregates and processes it, giving the $[Q_{(1,2)}, D_{(1,2)}]$ values. Finally, the root sensor ($S_{1, 1}$) gathers the statistical information from its immediately lower levels, processes it and generates the last statistics to be observed for anomaly detection. This final monitoring task is commonly carried out by a security analyst, who determines the presence of an anomaly when certain control limits are exceeded by the statistics \cite{camacho_pca-based_2016}.

Once the anomaly is detected, a deeper inspection should be done to determine, for example, where the anomaly comes from and the its reason. This is the diagnosis procedure, which is represented in Figure \ref{fig:working_together} with solid green and dashed brown lines, that is, the command and response involved actions, respectively. In this example, the anomaly comes from $S_{1,3}$, which is in charge of monitoring firewall traffic logs. First, a command message $[C_{m}]$ to find the anomaly origin is sent. How this message should be routed across the multilevel scenario is defined in the DRT, which maps RIS and LIS data sources known by each MSNM-Sensor and the observation timestamps. To determine which data source motivates the anomaly at a certain timestamp, a diagnosis algorithm is invoked. As mentioned before, oMEDA algorithm is currently used, though some other methods could easily be integrated. This procedure is repeated until the data source origin is found. Shortly thereafter, the involved piece of information (raw) falling into the observation period of time is returned to the root sensor to be analyzed in detail. A new message called response $[R_{m}]$ allows this operation.

\begin{figure*}[t!]
	\begin{centering}
		\includegraphics[width=0.7\textwidth]{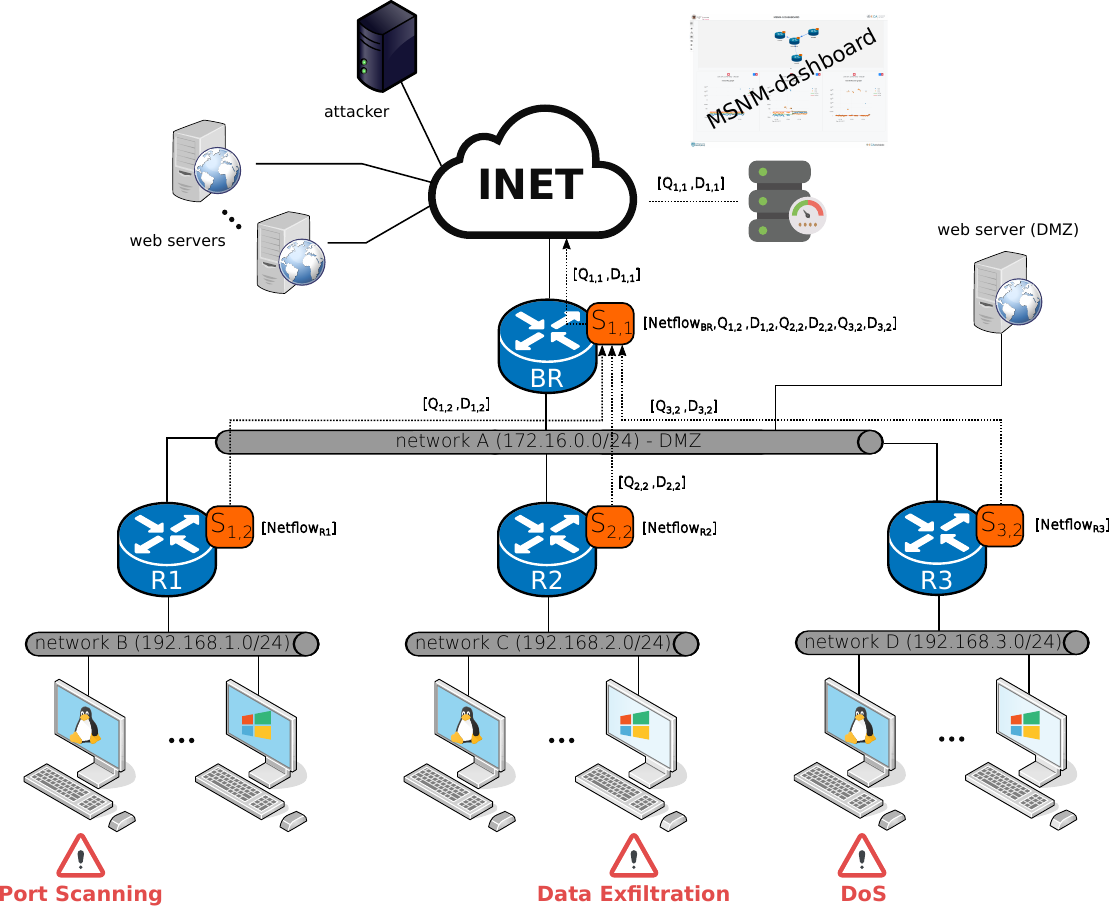}
		\par
	\end{centering}
	\caption{Experimental scenario. Four MSNM-Sensors ($S_{1,2}, S_{2,2}, S_{2,3}, S_{1,1}$) are deployed at each router. Each sensor monitors a NetFlow local data source generated from each router, while $S_{1,1}$ also aggregates all the statistics generated by the lower-level routers.}
	\label{fig:virtual_scenario}
	
\end{figure*}

\section{Practical Application of MSNM-Sensor for Monitoring and Attack Detection in Complex Network Environments}
\label{sec:experimentation}
\subsection{Experimental Environment}
To evaluate the performance of IDS-based systems, ready-to-use datasets are widely used. Consequently, choosing one or another dataset is a very relevant decision with considerable consequences, not only in obtaining results but also in establishing the validity of the conclusions the authors claim. In this way, Maci\'a-Fern\'andez \textit{et al.} \cite{macia-fernandez_ugr16_2018} build a recent and real network dataset that copes with the main drawbacks found in the most commonly used ones in the specialized literature. Nevertheless, not all of them are valid for use in certain application scenarios, because there are differences between the environment where the dataset was built and the one where the IDS solution will be deployed. In this manner, ready-to-use solutions for real-time anomaly detection are recommended. These types of approaches could eliminate the need to use previously gathered datasets, which are, on the other hand, very difficult to build.

%\textcolor{red}{[Hablarlo con Gabri acerca de como estaba montado y lo que yo extraje sobre todo en el número de máquinas windows, linux, 100 webservers, spoofing de IPs, etc] [Tengo que redactarlo de nuevo y mas breve para que no se parezca tanto a WIFS'16]}

To validate the monitoring and anomaly detection performance of the MSNM-Sensor in complex systems, we spread several sensors over a controlled scenario with virtual machines running in a cluster. This scenario has been previously devised to theoretically prove the hypothesis of the application of MSNM for hierarchical systems \cite{macia-fernandez_hierarchical_2016}. The key characteristics required are introduced in the following section. Interested readers can obtain more details in the mentioned reference. 

The complete scenario with the different machines is depicted in Figure \ref{fig:virtual_scenario}. This environment simulates a typical network architecture of a corporation so we can observe several subnetworks, network devices and end-devices. For instance, a DMZ is located in the inner network, separated from the outside world (Internet) with a BR (Border Router) and departmental networks in turn delimited by the corresponding routers (R1, R2, and R3).
%We can see a corporate network infrastructure connected to a simulated public Internet through a border router, BR. In the external part, we configure 100 virtual web servers and a machine where an attacker (a virtual machine with Kali2.0 distribution) is located. In what follows we describe the attacks that are carried out. Inside the network, there is a DMZ (network 172.16.0.0/24) with a single private web server, and three different departmental networks connected to the DMZ by routers R1, R2 and R3, respectively. Every departmental network is composed of \textcolor{red}{30 machines} (Windows-XP and Linux-Ubuntu-12.04).

In this scenario, we have two types of network traffic: normal and malicious. Normal traffic comprises all HTTP communications from all departmental hosts requesting HTTP resources allocated at the several web servers placed in the Internet and DMZ. As shown in Figure \ref{fig:virtual_scenario}, the BR is aware of all the incoming traffic from and outgoing traffic to the Internet. On the other hand, Rx routers, with $x={1,2,3}$, observe the corresponding portion of the previous HTTP traffic, which is generated by the hosts in their networks. Additionally, departmental hosts request HTTP resources to the web server in the DMZ.

On the other hand, the malicious traffic is generated from different locations in the predefined architecture, simulating very well-known and state-of-the-art attacks. These are 1) DoS (high and low rate); 2) port scanning, a relevant step in the recognition phase in a penetration testing procedure; and 3) data exfiltration for privacy violation purposes.

We run our scenario during a period of 25 hours. In the first 23.5 hours, only normal traffic is generated. During the last hour and a half, the attacks previously described are generated sequentially and do not overlap in time, with 5 minutes given for each one: high-rate DoS, low-rate DoS, scanning attack and data exfiltration.

%\begin{figure*}[t!]
%	\begin{centering}
%		\subfigure[]{
%			\includegraphics[scale=0.220] {img/borderRouter_roc}
%			\label{fig:roc:a}
%		}
%		\subfigure[]{
%			\includegraphics[scale=0.220] {img/routerR1_roc}
%			\label{fig:roc:b}
%		}
%		\subfigure[]{
%			\includegraphics[scale=0.220] {img/routerR2_roc}
%			\label{fig:roc:c}
%		}
%		\subfigure[]{
%			\includegraphics[scale=0.220] {img/routerR3_roc}
%			\label{fig:roc:d}
%		}
%		\caption{Performance ROC curves for the experiment attack period (last 1.5 hours) at  BR \subref{fig:roc:a}, R1 \subref{fig:roc:b}, R2 \subref{fig:roc:c} and R3 \subref{fig:roc:d}}
%		\label{fig:roc}
%	\end{centering}
%\end{figure*}

\begin{figure*}[t!]
	\begin{centering}
		\subfigure[]{
			\includegraphics[scale=0.220] {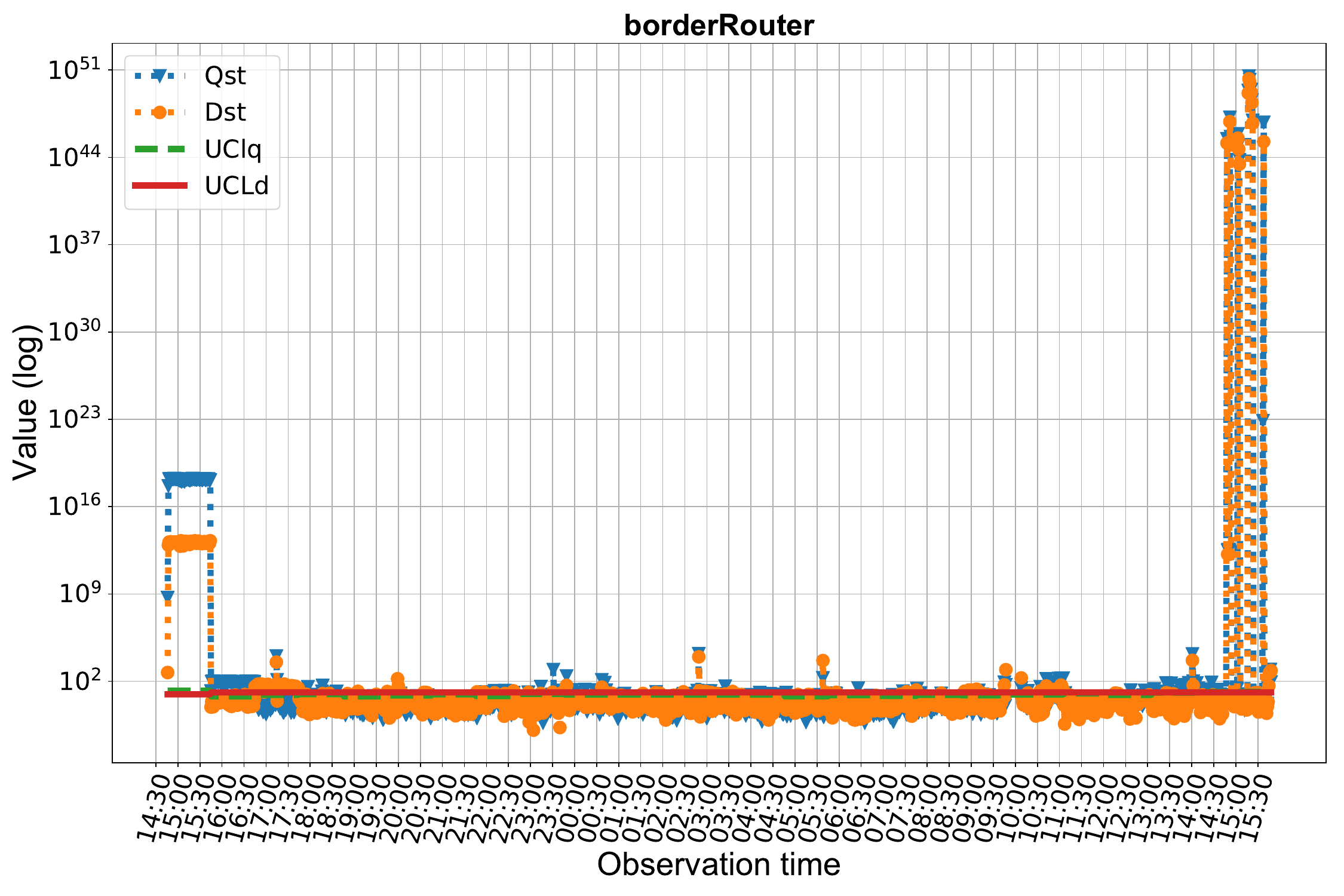}
			\label{fig:border_router_all_q_d:a}
		}
		\subfigure[]{
			\includegraphics[scale=0.220] {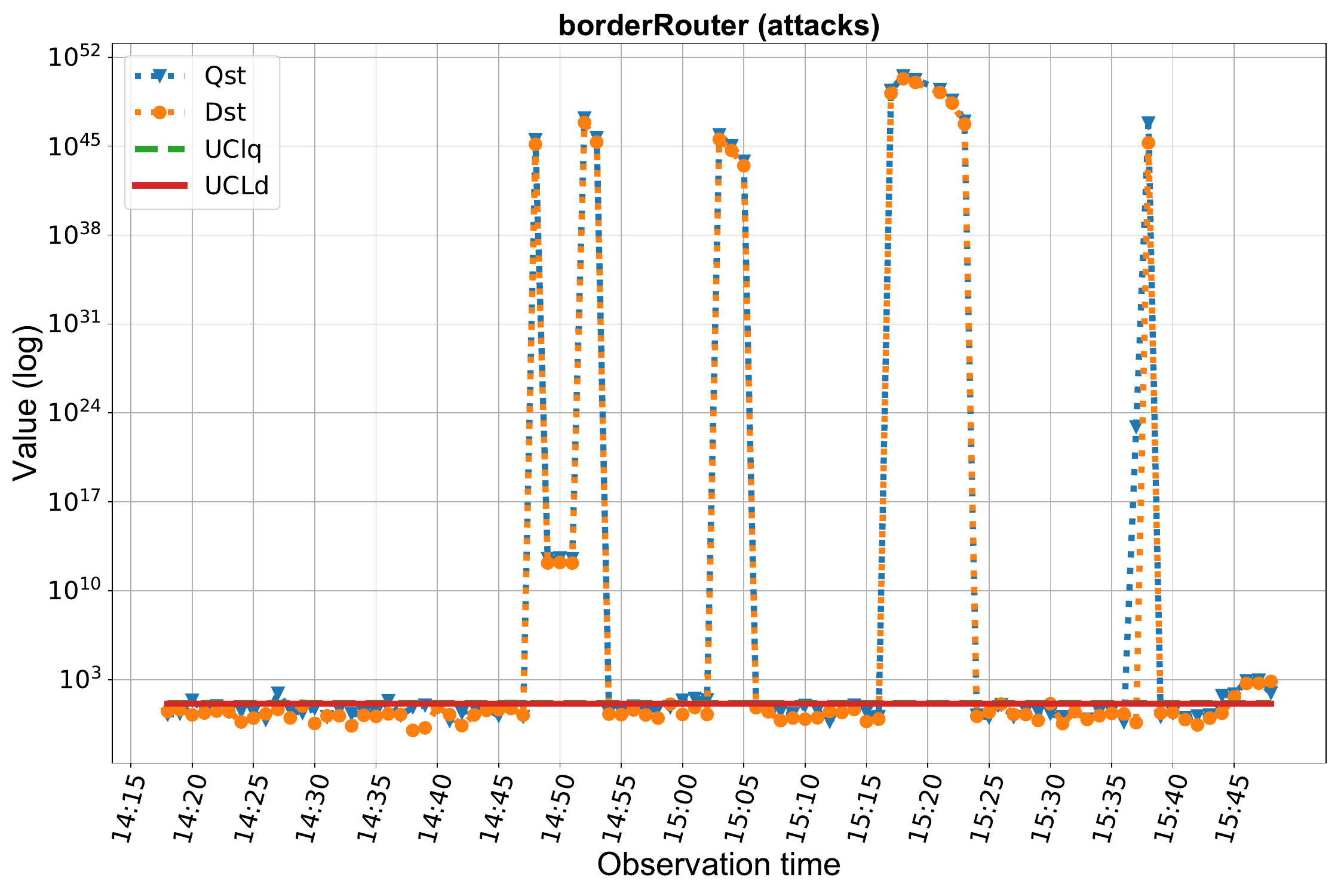}
			\label{fig:border_router_all_q_d:b}
		}
		\caption{$Q$-st and $D$-st statistics evolution for the entire experiment (25 hours) \subref{fig:border_router_all_q_d:a} and for the attack period (last 1.5 hours) \subref{fig:border_router_all_q_d:b}.}
		\label{fig:border_router_all_q_d}
	\end{centering}
\end{figure*}

The different routers in the network (R1, R2, R3 and BR) are equipped with NetFlow inspectors that generate NetFlow v5 information. These data are afterwards consumed in real time by the corresponding MSNM-Sensors, which are deployed in the mentioned network devices. These sensors are $S_{1,2},S_{2,2},S_{2,3}$ and $S_{1,1}$, which are represented by orange-colored boxes in Figure \ref{fig:virtual_scenario}. All of them consider only a local data source: the generated information of the corresponding NetFlow inspectors. However, only $S_{1,1}$ is in charge of aggregating the monitoring information in the form of statistics coming from the sensors in the network lower level. Every minute, a new observation is gathered by each sensor, meaning that two statistics are generated every minute.

\subsection{Experimental Results}
As we stated in previous sections, a key characteristic of the MSNM-Sensor is its applicability in real and complex environments monitoring a large variety of devices and communications. The MSNM-Sensor system is able to detect abnormal behaviors in the whole system considering only a couple of monitoring statistics. \footnote{To reproduce the obtained results, related data and scripts are available at the official GitHub repository.}

%[Introducir T-score y formula para computo de ROCs]

%In Figure \ref{fig:roc} we can see the MSNM-S performance for each of the deployed sensors at BR \ref{fig:roc:a}, R1 \ref{fig:roc:a}, R2 \ref{fig:roc:a} y R3 \ref{fig:roc:a}.

\begin{figure}[t!]
	\begin{centering}
		\includegraphics[scale=0.225]{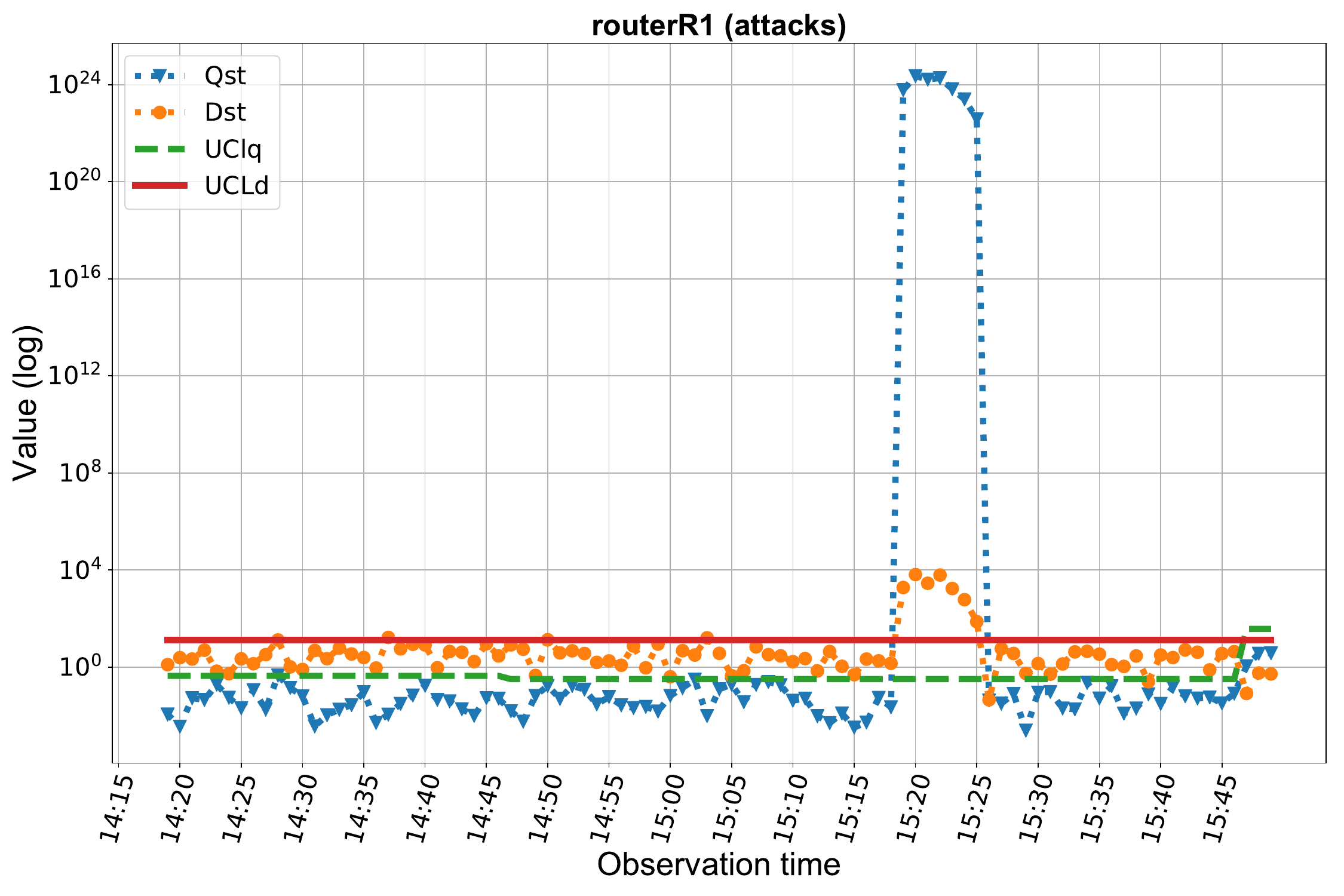}
		\par
	\end{centering}
	\caption{$Q$-st and $D$-st statistics evolution for the attack time interval in R1. It is clearly shown when the port scanning attack is taking place. }
	\label{fig:routerR1_anomaly_q_d}
	
\end{figure}

\begin{figure}
	\begin{centering}
		\includegraphics[scale=0.225]{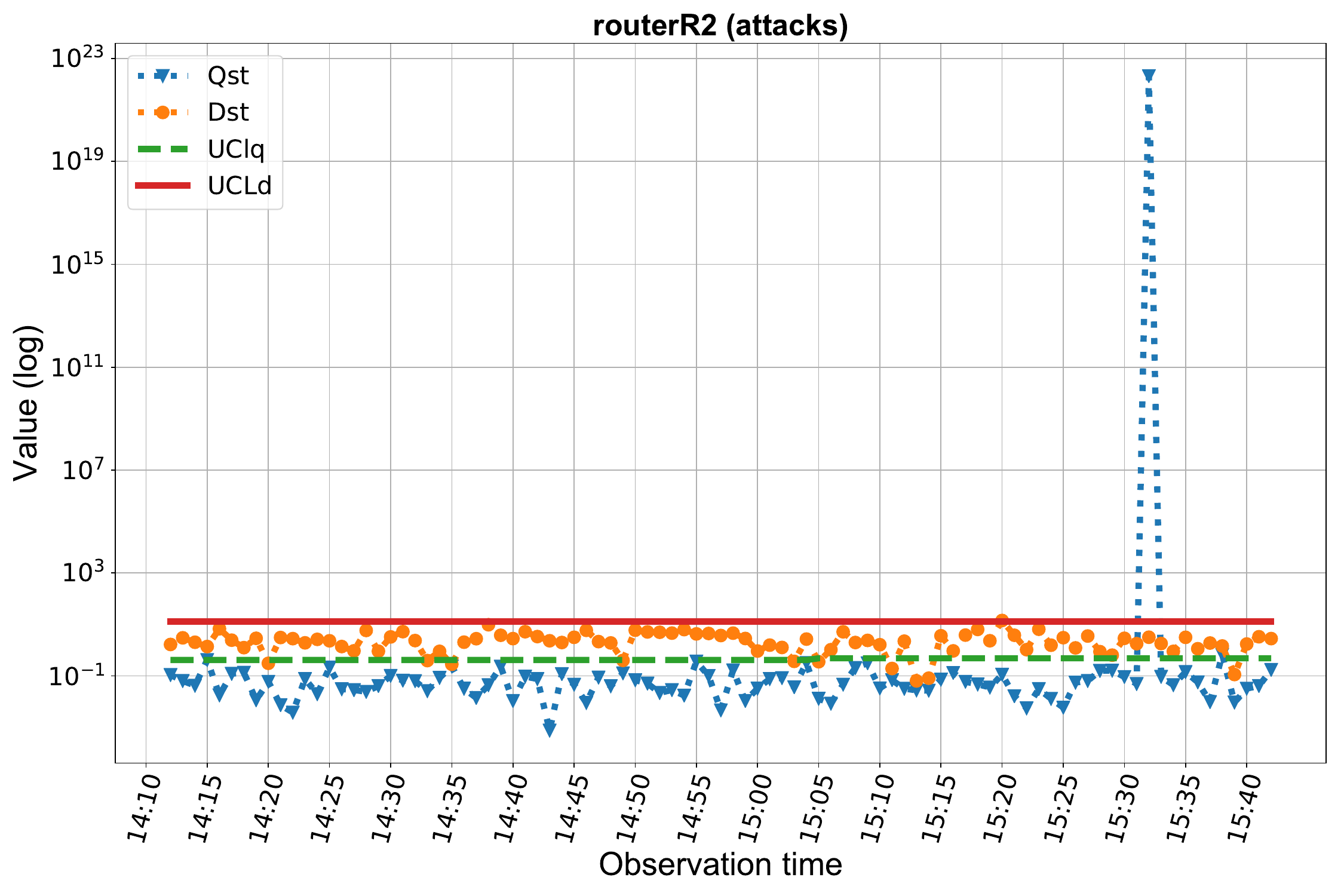}
		\par
	\end{centering}
	\caption{$Q$-st and $D$-st statistics evolution for the attack time interval in R2. It is clearly shown when the data exfiltration attack is taking place.}
	\label{fig:routerR2_anomaly_q_d}
	
\end{figure}

\begin{figure}
	\begin{centering}
		\includegraphics[scale=0.225]{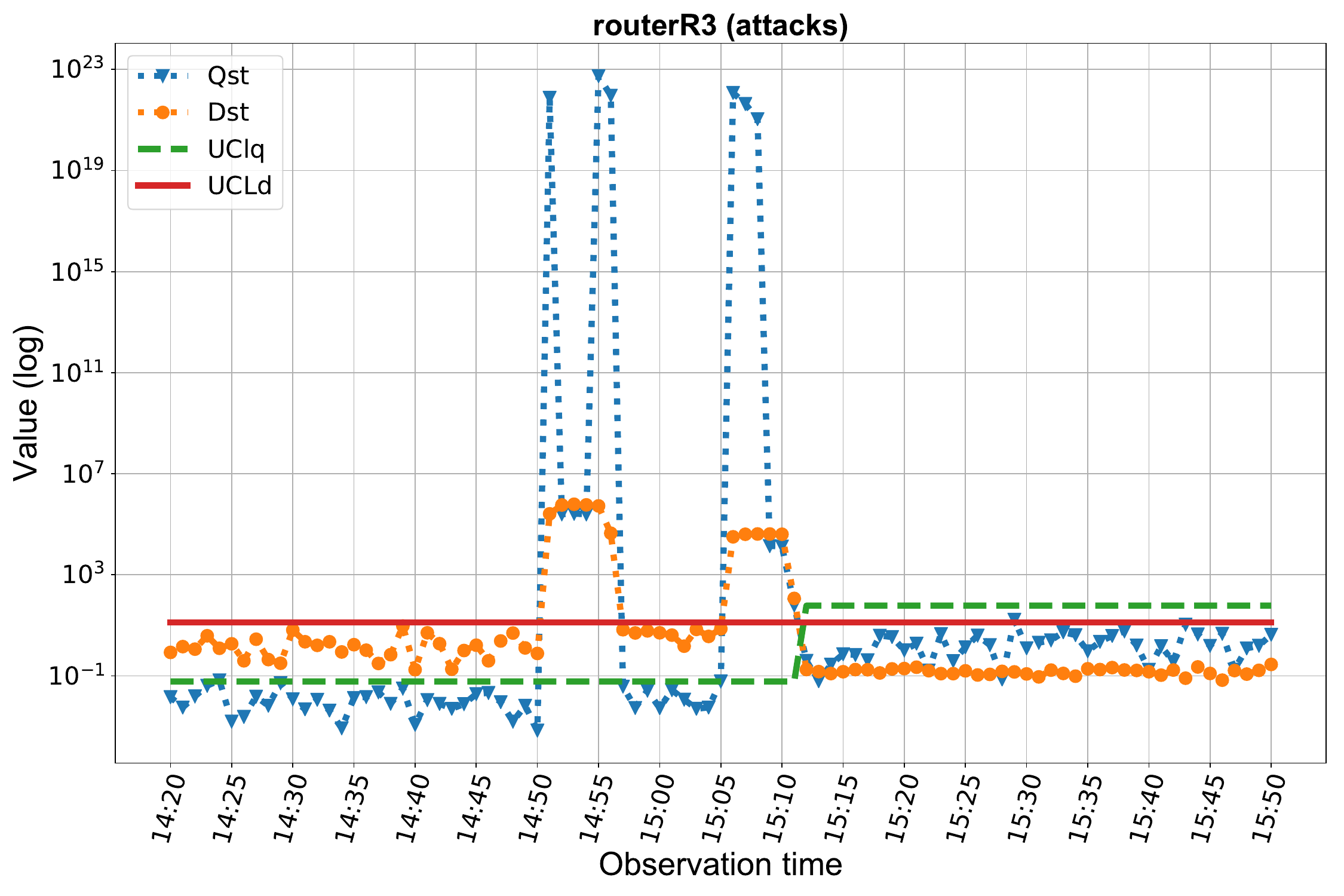}
		\par
	\end{centering}
	\caption{$Q$-st and $D$-st statistics evolution for the attack time interval in R3. It is clearly shown when the DoS attacks are taking place.}
	\label{fig:routerR3_anomaly_q_d}
	
\end{figure}

\begin{figure*}[t]
	\begin{centering}
		\includegraphics[width=0.925\textwidth]{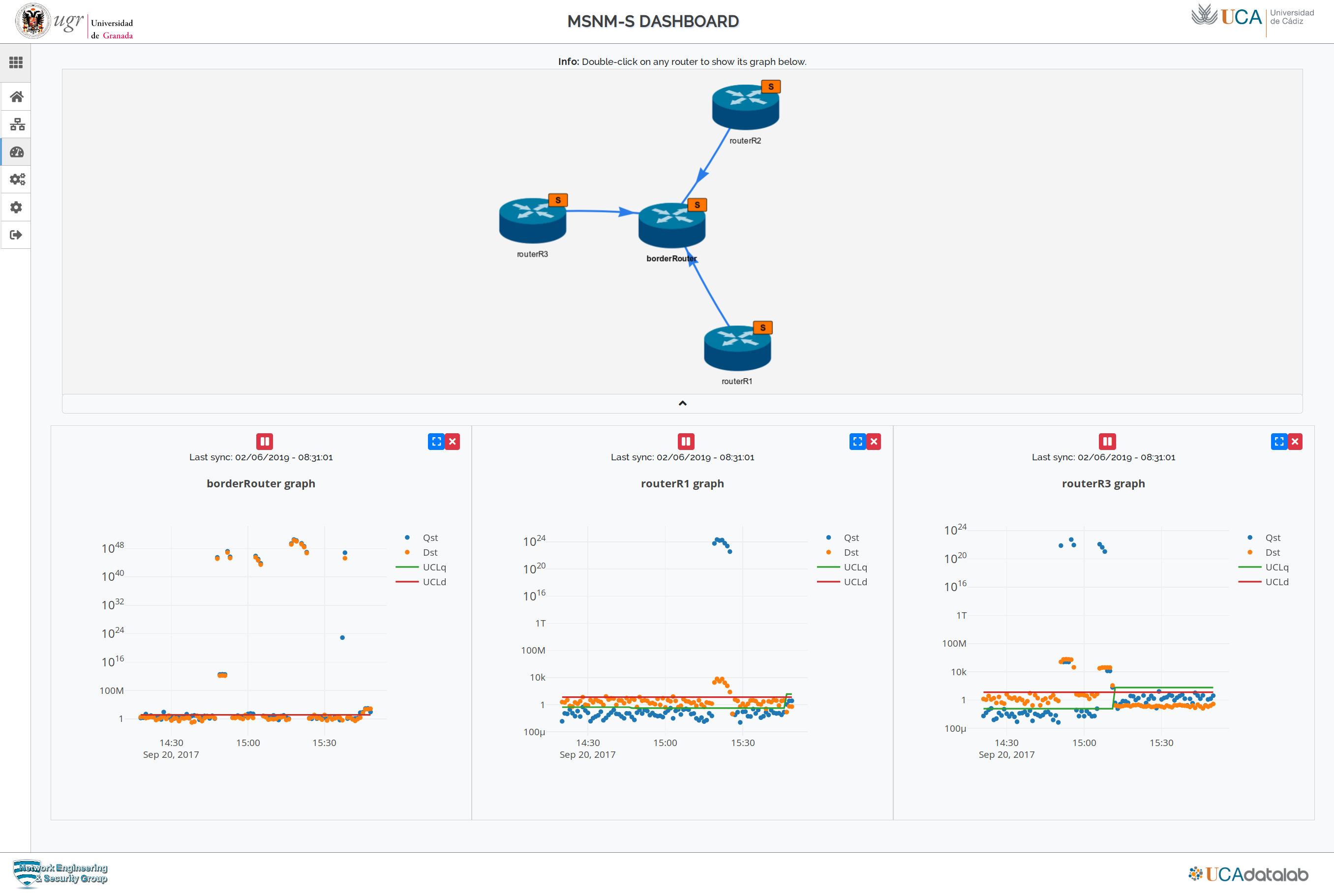}
		\par
	\end{centering}
	\caption{MSNM-Sensor dashboard snapshot of the monitoring view. The logical connections of the sensor (orange boxes) are shown in the upper section of the figure, while the monitoring graphs are in the bottom part.}
	\label{fig:dashboard_snapshot_all}
	
\end{figure*}

Figure \ref{fig:border_router_all_q_d} shows the $Q$-st (blue inverted triangles) and $D$-st (orange filled circles) statistics evolution with time obtained from the BR sensor. These kinds of graphics are the so-called monitoring plots. In addition, upper control limits are also shown for the $Q$-st (UCLq), represented by a green dashed line, and for the $D$-st (UCLd), represented by a red continuous line. Subfigure \ref{fig:border_router_all_q_d:a} shows the statistics values for the total duration of the experiment, while Subfigure \ref{fig:border_router_all_q_d:b} shows the last 1.5 hours, when the attacks were launched. In the first subfigure, we can discern between three different intervals. In the first one, we experience a high false positive rate, which is caused by the initial random calibration of the sensor. Each sensor uses the EWMA (Exponentially Weighted Moving Average) approach to dynamically calibrate the sensors every 60 minutes \cite{camacho_visualizing_2014}. The second one covers almost all the experimental time, where we can see the effectiveness of the dynamic adaptation of the sensor to the environment because the computed values are below the control limits mainly throughout the experimental time. Finally, the third one shows a clear deviation in the normal behavior of the statistics and the system in general. In this time period, the control limits are undoubtedly, exceeded indicating that something anomalous is occurring.

Apart from the previous global results, it is worth paying special attention to the attack period. In this way, Subfigure \ref{fig:border_router_all_q_d:b} shows clear deviations of the statistics values when the attacks are taking place. From DoS (high and low rate) to data exfiltration, the MSNM-Sensor sensor approach is able to detect anomalous behaviors coming from different parts of the whole systems by taking into account only two simple numerical values. Moreover, we can distinguish where the anomaly is coming from by inspecting similar monitoring graphics, but in this case, they are computed from each of the involved routers. This process can be viewed in Figures \ref{fig:routerR1_anomaly_q_d}, \ref{fig:routerR2_anomaly_q_d} and \ref{fig:routerR3_anomaly_q_d} for the sensors deployed at R1, R2 and R3, respectively. For instance, from the inspection of the R3 monitoring graph, we can conclude that the attack originated somewhere in the R3 network, similarly for R1 and R2 for the port scanning and data exfiltration attacks, respectively. Additionally, we can see the dynamic adaptation on the sensor to the changes in the environment. It is more evident in the R3 monitoring graphic (see Figure \ref{fig:routerR3_anomaly_q_d}), where the UCLq limit adapts its value to cover new trends in $Q$-st.  

Although the previous figures are included in the text to enhance the reading, the MSNM-Sensor application comes with a specific and interactive dashboard. Thanks to this dashboard, the security analyst or the application operator can, in real time, access the previous information and the logical connections of the sensors. A snapshot of the monitoring main section of the tool is shown in Figure \ref{fig:dashboard_snapshot_all}. The upper part shows the logical connections created among the sensors. The operator can see the direction of the monitoring flow. In this specific scenario, it is clearly shown how the sensors in routers R1, R2 and R3 are configured to send their monitoring information to the BR. In addition, the monitoring graphs will appear in the bottom part by clicking on a specific sensor. Among other actions, the operator can interact with these graphs to pause/play the updating procedure for a better inspection. The dashboard is needed for whatever monitoring system that allows the operator to reduce the response/reaction time when an attack takes place.

\section{Conclusions \& Future work}
\label{sec:conclusions}
This work introduces the MSNM-Sensor (Multivariate Statistical Network Monitoring Sensor), a practical and ready-to-use tool for monitoring and detecting security events in complex systems and network environments. The MSNM-S relies on the implementation of the MSNM methodology which allows it address new technological challenges arising from all-connected scenarios, \textit{e.g.}, smart cities and IoT ecosystems, where a large number of devices share information. 

Inherited from the MNSM methodology, MSNM-Sensor reduces and efficiently handles the monitoring information, maintaining its detection performance. On the other hand, existing IDS or SIEM solutions consider raw data, thus adding extra monitoring traffic overhead. Moreover, MSNM-Sensor inherently adds privacy in monitoring communications since no sensible or raw data is sent to the central station just the monitoring statistics instead.

The MSNM-Sensor uses lightweight algebraic and statistics operations that allow it to be embedded in less powerful devices, \textit{e.g.}, wearable devices, environmental sensors and IoT devices in general.

Last but not least, the MSNM-Sensor operates in real time, and it is adaptable to changes in the environment where it is deployed without any previous training phases.

Although the MSNM-Sensor has been tested successfully with promising results in hierarchical network systems and architectures to monitor and detect well-known network attacks, the MSNM-Sensor capabilities should be tested in real and complex scenarios, \textit{e.g.}, IoT ecosystems or those found in smart cities. Finally, the anomaly diagnosis operation is still in development; thus, we will focus our future work on this topic.

%%% inclusión de referncias
\bibliographystyle{IEEEtran}
\bibliography{msnmsensor}

% Generated by IEEEtran.bst, version: 1.13 (2008/09/30)
\begin{thebibliography}{10}
\providecommand{\url}[1]{#1}
\csname url@samestyle\endcsname
\providecommand{\newblock}{\relax}
\providecommand{\bibinfo}[2]{#2}
\providecommand{\BIBentrySTDinterwordspacing}{\spaceskip=0pt\relax}
\providecommand{\BIBentryALTinterwordstretchfactor}{4}
\providecommand{\BIBentryALTinterwordspacing}{\spaceskip=\fontdimen2\font plus
\BIBentryALTinterwordstretchfactor\fontdimen3\font minus
  \fontdimen4\font\relax}
\providecommand{\BIBforeignlanguage}[2]{{%
\expandafter\ifx\csname l@#1\endcsname\relax
\typeout{** WARNING: IEEEtran.bst: No hyphenation pattern has been}%
\typeout{** loaded for the language `#1'. Using the pattern for}%
\typeout{** the default language instead.}%
\else
\language=\csname l@#1\endcsname
\fi
#2}}
\providecommand{\BIBdecl}{\relax}
\BIBdecl

\bibitem{iot_devices_outdated}
A.~Nordrum, ``Popular internet of things forecast of 50 billion devices by 2020
  is outdated.'' [Online; Accessed 30-September-2019]
  \url{https://bit.ly/2kVkk9A}.

\bibitem{m2m_connections}
{Cisco Systems}, ``Cisco visual networking index: Global mobile data traffic
  forecast update, 2017-2022 white paper,'' [Online; Accessed
  30-September-2019] \url{https://bit.ly/2TYO1DG}.

\bibitem{enisa_etl_2017}
{ENISA}, ``{ENISA} {Threat} {Landscape} {Report} 2017,'' [Online; Accessed
  22-September-2019]
  \url{https://www.enisa.europa.eu/publications/enisa-threat-landscape-report-2017}.

\bibitem{chaabouni_network_2019}
N.~Chaabouni, M.~Mosbah, A.~Zemmari, C.~Sauvignac, and P.~Faruki, ``Network
  {Intrusion} {Detection} for {IoT} {Security} {Based} on {Learning}
  {Techniques},'' \emph{IEEE Communications Surveys Tutorials}, vol.~21, no.~3,
  pp. 2671--2701, 2019.

\bibitem{enisa_etl_2018}
{ENISA}, ``{ENISA} {Threat} {Landscape} {Report} 2018,'' [Online; Accessed
  22-September-2019]
  \url{https://www.enisa.europa.eu/publications/enisa-threat-landscape-report-2018}.

\bibitem{camacho_tackling_2014}
J.~Camacho, G.~Maci\'a-Fern\'andez, J.~D\'iaz-Verdejo, and P.~Garc\'ia-Teodoro,
  ``Tackling the {Big} {Data} 4 vs for anomaly detection,'' in \emph{2014
  {IEEE} {Conference} on {Computer} {Communications} {Workshops} ({INFOCOM}
  {WKSHPS})}, April 2014, pp. 500--505.

\bibitem{snort_ids}
{Cisco Systems}, ``Snort {IDS},'' [Online; Accessed 30-September-2019]
  \url{https://www.snort.org/}.

\bibitem{bro_ids}
I.~Johanna~Amann, ``Zeek (bro) {IDS},'' [Online; Accessed 30-September-2019]
  \url{https://www.zeek.org/}.

\bibitem{splunk_siem}
Splunk, ``Splunk,'' [Online; Accessed 30-September-2019]
  \url{https://www.splunk.com}.

\bibitem{alient_siem}
A.~Vault, ``Alient vault,'' [Online; Accessed 30-September-2019]
  \url{https://www.alienvault.com/}.

\bibitem{camacho_pca-based_2016}
J.~Camacho, A.~P\'erez-Villegas, P.~Garc\'ia-Teodoro, and
  G.~Maci\'a-Fern\'andez, ``{PCA}-based multivariate statistical network
  monitoring for anomaly detection,'' \emph{Computers \& Security}, vol.~59,
  pp. 118--137, June 2016.

\bibitem{camacho_traffic_2017}
J.~Camacho, P.~García-Teodoro, and G.~Maciá-Fernández, ``Traffic
  {Monitoring} and {Diagnosis} with {Multivariate} {Statistical} {Network}
  {Monitoring}: {A} {Case} {Study},'' in \emph{2017 {IEEE} {Security} and
  {Privacy} {Workshops} ({SPW})}, May 2017, pp. 241--246.

\bibitem{camacho_semi-supervised_2019}
J.~Camacho, G.~Maci\'a-Fern\'andez, N.~M. Fuentes-Garc\'ia, and E.~Saccenti,
  ``Semi-supervised {Multivariate} {Statistical} {Network} {Monitoring} for
  {Learning} {Security} {Threats},'' \emph{IEEE Transactions on Information
  Forensics and Security}, pp. 1--1, 2019.

\bibitem{camacho_multivariate_2019}
J.~Camacho, J.~M. García-Giménez, N.~M. Fuentes-García, and
  G.~Maciá-Fernández, ``\BIBforeignlanguage{en}{Multivariate {Big} {Data}
  {Analysis} for intrusion detection: 5 steps from the haystack to the
  needle},'' \emph{\BIBforeignlanguage{en}{Computers \& Security}}, vol.~87,
  pp. 1--11, November 2019.

\bibitem{kanaoka_multivariate_2003}
A.~Kanaoka and E.~Okamoto, ``Multivariate statistical analysis of network
  traffic for intrusion detection,'' in \emph{14th {International} {Workshop}
  on {Database} and {Expert} {Systems} {Applications}, 2003. {Proceedings}.},
  September 2003, pp. 472--476, iSSN: 1529-4188.

\bibitem{lakhina_diagnosing_2004}
A.~Lakhina, M.~Crovella, and C.~Diot, ``Diagnosing network-wide traffic
  anomalies,'' in \emph{Proceedings of the 2004 conference on {Applications},
  technologies, architectures, and protocols for computer communications}, ser.
  {SIGCOMM} '04.\hskip 1em plus 0.5em minus 0.4em\relax Portland, Oregon, USA:
  Association for Computing Machinery, August 2004, pp. 219--230.

\bibitem{callegari_improving_2014}
C.~Callegari, L.~Gazzarrini, S.~Giordano, M.~Pagano, and T.~Pepe,
  ``\BIBforeignlanguage{en}{Improving {PCA}-based anomaly detection by using
  multiple time scale analysis and {Kullback}–{Leibler} divergence},''
  \emph{\BIBforeignlanguage{en}{International Journal of Communication
  Systems}}, vol.~27, no.~10, pp. 1731--1751, 2014.

\bibitem{delimargas_evaluating_2014}
A.~Delimargas, E.~Skevakis, H.~Halabian, I.~Lambadaris, N.~Seddigh, B.~Nandy,
  and R.~Makkar, ``Evaluating a modified {PCA} approach on network anomaly
  detection,'' in \emph{2014 {International} {Conference} on {Next}
  {Generation} {Networks} and {Services} ({NGNS})}, May 2014, pp. 124--131.

\bibitem{magan-carrion_multivariate_2015}
R.~Mag\'an-Carri\'on, J.~Camacho, and P.~Garc\'ia-Teodoro,
  ``\BIBforeignlanguage{en}{Multivariate {Statistical} {Approach} for {Anomaly}
  {Detection} and {Lost} {Data} {Recovery} in {Wireless} {Sensor}
  {Networks}},'' \emph{\BIBforeignlanguage{en}{International Journal of
  Distributed Sensor Networks}}, vol.~11, no.~6, pp. 1--20, June 2015.

\bibitem{faac_github}
A.~P\'erez-Villegas, J.~Garc\'ia-Jim\'enez, and J.~Camacho, ``{FaaC}
  (feature-as-a-counter) parser - {Github},'' [Online; Accessed
  30-September-2019] \url{https://github.com/josecamachop/FCParser}.

\bibitem{jackson_control_1979}
J.~E. Jackson and G.~S. Mudholkar, ``Control {Procedures} for {Residuals}
  {Associated} with {Principal} {Component} {Analysis},'' \emph{Technometrics},
  vol.~21, no.~3, pp. 341--349, 1979.

\bibitem{hotelling1947}
H.~Hotelling, \emph{Multivariate Quality Control. Techniques of Statistical
  Analysis}.\hskip 1em plus 0.5em minus 0.4em\relax MacGraw-Hill, 1947.

\bibitem{nomikos_multivariate_1995}
P.~Nomikos and J.~F. MacGregor, ``Multivariate {SPC} {Charts} for {Monitoring}
  {Batch} {Processes},'' \emph{Technometrics}, vol.~37, no.~1, pp. 41--59,
  February 1995.

\bibitem{camacho_observation-based_2011}
J.~Camacho, ``\BIBforeignlanguage{en}{Observation-based missing data methods
  for exploratory data analysis to unveil the connection between observations
  and variables in latent subspace models},''
  \emph{\BIBforeignlanguage{en}{Journal of Chemometrics}}, vol.~25, no.~11, pp.
  592--600, November 2011.

\bibitem{nmfuentes_2018}
M.~Fuentes-Garc\'ia, G.~Maci\'a-Fern\'andez, and J.~Camacho, ``Evaluation of
  diagnosis methods in {PCA}-based {Multivariate} {Statistical} {Process}
  {Control},'' \emph{Chemometrics and Intelligent Laboratory Systems}, vol.
  172, pp. 194--210, January 2018.

\bibitem{bhuyan_network_2014}
M.~H. Bhuyan, D.~K. Bhattacharyya, and J.~K. Kalita, ``Network {Anomaly}
  {Detection}: {Methods}, {Systems} and {Tools},'' \emph{IEEE Communications
  Surveys Tutorials}, vol.~16, no.~1, pp. 303--336, 2014.

\bibitem{li_machine_2019}
Z.~Li, A.~L.~G. Rios, G.~Xu, and L.~Trajković, ``Machine {Learning}
  {Techniques} for {Classifying} {Network} {Anomalies} and {Intrusions},'' in
  \emph{2019 {IEEE} {International} {Symposium} on {Circuits} and {Systems}
  ({ISCAS})}, May 2019, pp. 1--5.

\bibitem{macia-fernandez_ugr16_2018}
G.~Maci\'a-Fern\'andez, J.~Camacho, R.~Mag\'an-Carri\'on, P.~Garc\'ia-Teodoro,
  and R.~Ther\'on, ``{UGR}‘16: {A} new dataset for the evaluation of
  cyclostationarity-based network {IDSs},'' \emph{Computers \& Security},
  vol.~73, pp. 411--424, March 2018.

\bibitem{camacho_visualizing_2014}
J.~Camacho, ``Visualizing {Big} data with {Compressed} {Score} {Plots}:
  {Approach} and research challenges,'' \emph{Chemometrics and Intelligent
  Laboratory Systems}, vol. 135, pp. 110--125, July 2014.

\bibitem{msnm-sensor_github}
R.~Mag\'an-Carri\'on, J.~Camacho, and G.~Maci\'a-Fern\'andez, ``{MSNM-S}:
  {Multivariate} {Statistical} {Network} {Monitoring}-{Sensor} - {Github},''
  [Online; Accessed 30-September-2019]
  \url{https://github.com/nesg-ugr/msnm-sensor}.

\bibitem{macia-fernandez_hierarchical_2016}
G.~Maci\'a-Fern\'andez, J.~Camacho, P.~Garc\'ia-Teodoro, and R.~A.
  Rodr\'iguez-G\'omez, ``Hierarchical {PCA}-based multivariate statistical
  network monitoring for anomaly detection,'' in \emph{2016 {IEEE}
  {International} {Workshop} on {Information} {Forensics} and {Security}
  ({WIFS})}, December 2016, pp. 1--6.

\end{thebibliography}

\end{document}